\documentclass[12pt,amsmath]{iopart}
\usepackage{graphicx}
\usepackage{amssymb,color}

\newcommand{\vc}{\mathbf}
\newcommand{\stres}{\ensuremath{\Pi}}
\newcommand{\ident}{\ensuremath{\mathcal{I}}}

\begin{document}

\title[Baalrud and Hegna Reply]{Reply to Comment on ``Kinetic Theory of the Presheath and the Bohm Criterion''}

\author{S D Baalrud}

\address{Theoretical Division, Los Alamos National Laboratory, Los Alamos, NM 87545, USA}
\address{Department of Physics and Astronomy, University of Iowa, Iowa City, IA, 52242, USA}

\author{C C Hegna}

\address{Department of Engineering Physics, University of Wisconsin-Madison, 1500 Engineering Drive, Madison, WI 53706, USA}

\begin{abstract}

In his comment, Riemann defends the conventional kinetic Bohm criterion on the basis that the underlying approximations become rigorous in the limit $\lambda_{D}/l \rightarrow 0$, where $\lambda_{D}$ is the Deybe length and $l$ is a collision length scale for the dominant collision process. Here, we expand on our previous arguments showing that the basic assumptions for the justified use of the conventional kinetic Bohm criterion are typically not satisfied in laboratory plasmas. We corroborate our argument with experimental data, as well as data from numerical simulations, showing that the conventional criterion is violated in common situations. In contrast, a formulation based on positive velocity moments of the kinetic equation provides a criterion that both agrees with the experimental data and reduces to the traditional expectations from fluid theory in the appropriate limit. 

\end{abstract}
\submitto{\PSST}
\maketitle


In his comment \cite{riem:12}, Riemann defends the conventional kinetic Bohm criterion (KBC) \cite{harr:59,riem:81,riem:91}
\begin{equation}
\frac{1}{M_i} \int d^3v \frac{f_i}{v_z^2} \leq - \frac{1}{m_e} \int d^3v \frac{1}{v_z} \frac{\partial f_e}{\partial v_z} \label{eq:kbc}
\end{equation}
using theoretical arguments based on the Vlasov equation and the assumption of a collisionless, or infinitely thin, sheath $\lambda_D/ l \rightarrow 0$. Here $\lambda_D$ is the Debye length, which is the characteristic scale of the sheath, and $l$ is the characteristic scale of the plasma (or presheath), which is typically the shortest ion collision mean free path. Such two-scale analyses have been very successful at modeling important quantities such as ion and electron density, electrostatic potential and current profiles in the plasma-boundary region when $\lambda_D/l \ll 1$ \cite{riem:81,riem:91,tonk:29,caru:62,fran:70,emme:80,biss:87,sche:88}. Some of these also provided models of the ion velocity distribution function (IVDF) \cite{riem:81,tonk:29,emme:80,biss:87,sche:88} that have been shown to qualitatively predict experimental~\cite{bach:95,clai:06} and numerical simulation data \cite{koch:89,proc:90,sher:01,robe:09}. However, there is one important difference; the IVDF is always empty for $v_z \geq 0$ at the sheath edge in the theories, but not in the data. Theoretically, this is a consequence of the infinitely thin sheath assumption (no ions can escape the sheath). Experimentally, the sheath has a finite thickness, and a small fraction can escape either from scattering or from ions born from ionization that have sufficient directed energy to escape the sheath. Although this fraction is small, it has major consequences for the applicability of Eq.~(\ref{eq:kbc}) because the left side diverges in this situation. In this Reply, we reiterate the main point of Ref.~\cite{baal:11}; Eq.~(\ref{eq:kbc}) is not valid for arbitrary distribution functions, including cases experimentally measured. Instead, an alternative KBC based on positive-exponent velocity moments should be used. 

Bohm's original criterion \cite{bohm:49}, $V_i \geq c_s$, was derived using two-fluid theory in the cold ion limit. The goal of a kinetic Bohm criterion is to generalize the fluid approach to account for arbitrary electron and ion distribution functions. Equation~(\ref{eq:kbc}) applies to only a subset of possible ion distributions because the derivation assumes $f_i (v_z = 0) = 0$. Although this condition is satisfied for theoretical IVDFs calculated in the limit $\lambda_D/l \rightarrow 0$, it is not typically satisfied experimentally because sheaths have a finite thickness.  Equation~(\ref{eq:kbc}) also assumes the distribution functions are Vlasov solutions, and utilizes properties such as $f_i$ is a function of $v_z^2$ only. However, non-Vlasov distribution functions are interesting because the presheath electric field always generates a current, implying that ions and electrons are not both Vlasov solutions in the same reference frame. One must be cognizant of the errors introduced when using non-Vlasov solutions in a theory based on the Vlasov equation. 

We are not aware of an experiment designed specifically to test Eq.~(\ref{eq:kbc}), but experimental and numerical data has appeared in the literature that appear to violate it. Several experimental papers have presented measurements of the IVDF through the presheath and sheath regions, including the sheath edge. The primary experimental technique has been laser induced fluorescence (LIF) \cite{bach:95,clai:06,goec:92,bach:93,carr:96,sade:97,oksu:01,lunt:08,clai:12,seve:03,lee:06,jaco:07,jaco:10}. This has been applied in a variety of situations including single ion species plasmas~\cite{bach:95,clai:06,goec:92,bach:93,carr:96,sade:97,oksu:01,lunt:08,clai:12,seve:03}, multiple ion species plasmas \cite{seve:03,lee:06}, and RF plasmas \cite{jaco:07,jaco:10}.  A variety of numerical simulation techniques have also been applied to calculate the IVDF in the plasma-boundary region~\cite{koch:89,sher:01,robe:09,tsus:98,milo:11}. IVDFs have also been measured near double layers, which are sheath-like structures where the Bohm criterion also applies, using LIF \cite{sun:05} and retarding field energy analyzers (RFEA) \cite{scie:10}, as well as calculation from particle-in-cell (PIC) simulations \cite{meig:05,baal:11b}. Each of these references provide example IVDFs with a finite number of particles for $v_z \geq 0$ at the sheath edge. 

\begin{figure}
\includegraphics[width=6.0in]{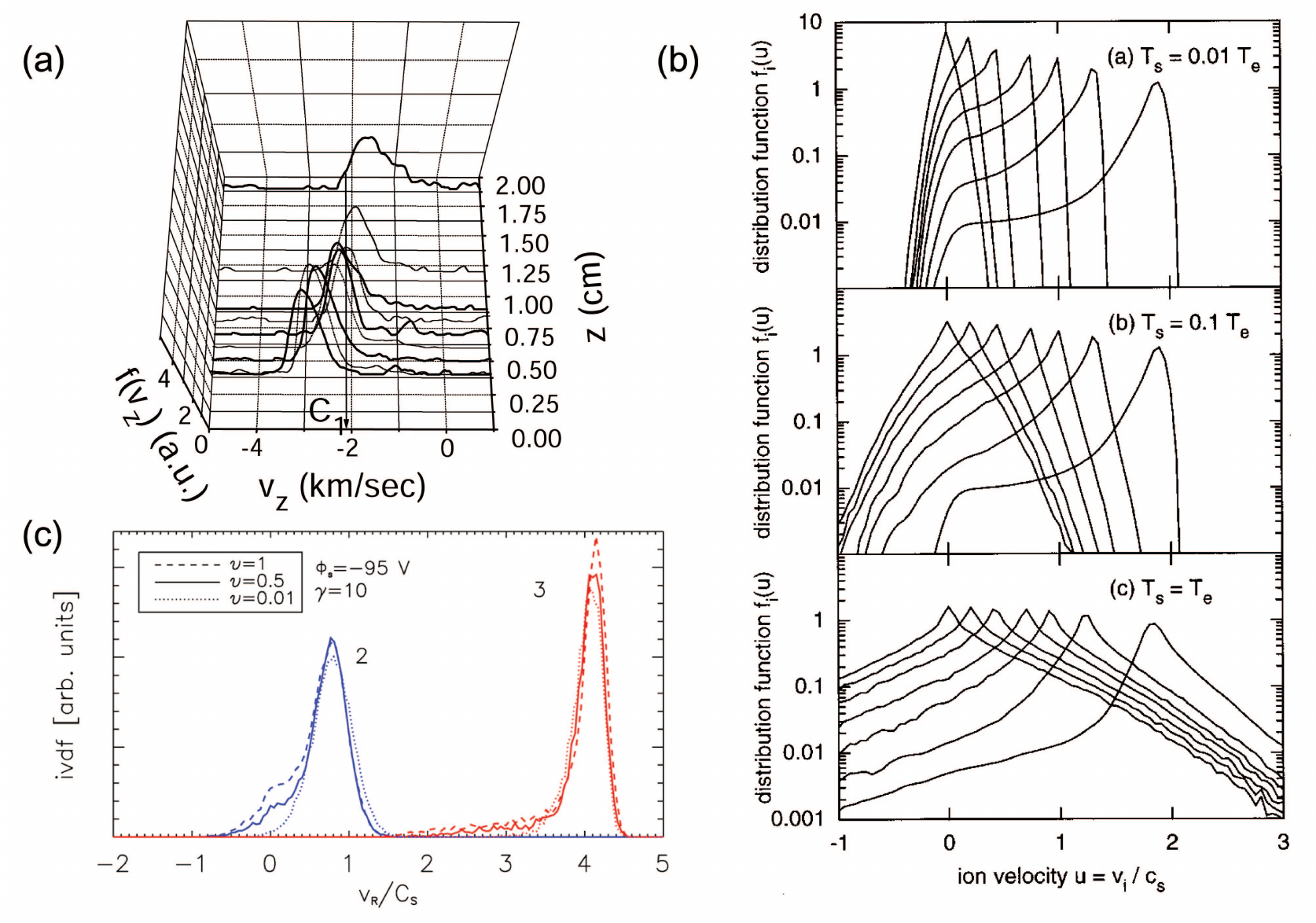}
\caption{ IVDFs in the plasma-boundary region from (a) LIF measurements from \cite{seve:03}, (b) Hybrid simulations from \cite{sher:01}, and (c) PIC simulations from \cite{milo:11}. Reprinted with permission from: (a) G.\ D.\ Severn, X.\ Wang, E.\ Ko and N.\ Hershkowitz, Phys.\ Rev.\ Lett.\ {\bf 90}, 145001 (2003). Copyright 2003, American Physical Society, (b) T.\ E.\ Sheridan, Phys.\ Plasmas {\bf 8}, 4240 (2001). Copyright 2001, American Institute of Physics, (c) W.\ J.\ Miloch, N.\ Gulbrandsen, L.\ N.\ Mishra and \AA.\ Fredriksen, Phys.\ Plasmas {\bf 18}, 083502 (2011). Copyright 2011, American Institute of Physics. }
\label{fg:data}
\end{figure}

Three of these examples are shown in Fig.~\ref{fg:data}. Panel (a) shows experimental LIF data of an argon plasma from \cite{seve:03}. Panel (b) shows data from a hybrid simulation using a PIC routine for ions, and assuming electrons obey the Boltzmann density relation \cite{sher:01}. Panel (c) shows data from a PIC simulation pushing both ions and electrons \cite{milo:11}. In each of these examples, there is a nonzero contribution to the IVDF for $v_z=0$ at the sheath edge.\footnote{It is hard to tell what the ion density is at $v_z=0$ for the LIF data in case (a) because the signal in this region of velocity space is within the noise. However, it is also hard to apply Eq.~(\ref{eq:kbc}) to this data because the $v_z^{-2}$ moment amplifies the noise. For an example measurement with a clearer signal at $v_z=0$ see Fig.~3 of \cite{lunt:08}. } Consequently, the left side of Eq.~(\ref{eq:kbc}) diverges and the resultant criteria is not valid. It is useful to note that although the $\lambda_D / l \rightarrow 0$ limit is used to identify the sheath edge in the conventional theory, the sheath edge is experimentally meaningful even when this quantity is finite, so long as it is small. A variety of techniques are availible for estimating the location of the sheath edge. In fact, for the three sets of data in Fig.~\ref{fg:data}, three different techniques have been used. The experiments in (a) used a technique based on emissive probe theory~\cite{wang:06}, estimating the sheath edge at $z = 6.0 \pm 0.5$ mm. The hybrid simulations in (b) used the location for divergence of $E$ from the cold Tonks-Langmuir problem, estimating the sheath edge at $x/L = 0.917$. The full PIC simulations in (c) used an estimate based on a collisionless approximation, taking the sheath edge to be where the electrostatic potential has dropped $T_e/2$ from the bulk plasma value; leading to an estimate of $9 \lambda_{D}$ from the boundary.  In each case a meaningful sheath edge can be identified, but Eq.~(\ref{eq:kbc}) does not provide a meaningful Bohm criterion since the left side diverges. 

Reference \cite{baal:11} considers a couple of possible reasons for this disagreement. One of these is related to a step in the derivation of Eq.~(\ref{eq:kbc}) where integration-by-parts is used (see section 2.3 of \cite{baal:11} and section 3c of \cite{fern:05}). Since this step is only valid for continuously differentiable functions, the restriction $f_i (v_z = 0) = 0$ must be enforced. Instead of applying the integration-by-parts, and thereby restricting the set of IVDFs the theory can be applied to, one may write the criterion in the more primitive form from before this step is taken 
\begin{equation}
\frac{1}{M_i} \int_{-\infty}^\infty d^3v\, \frac{1}{v_z} \frac{\partial f_i}{\partial v_z} \leq - \frac{1}{m_e} \int_{-\infty}^\infty d^3v \frac{1}{v_z} \frac{\partial f_e}{\partial v_z} .  \label{eq:kbcp}
\end{equation}
Equation~(\ref{eq:kbcp}) should apply to an expanded set of distribution functions compared to Eq.~(\ref{eq:kbc}). However, the derivation of Eq.~(\ref{eq:kbcp}) still relies on symmetry properties of Vlasov solutions, and the errors introduced by using non-Vlasov solutions in this equation may be significant. Although Eq.~(\ref{eq:kbcp}) arises as an intermediate step in the derivation of Eq.~(\ref{eq:kbc}), Riemann argues \cite{riem:12,riem:03,riem:06} that there is no useful sheath-related information in Eq.~(\ref{eq:kbcp}) because it is satisfied everywhere and that the integration-by-parts step, with the corresponding $f_i(v_z=0)=0$ restriction, is required to obtain a meaningful criterion. 

\begin{figure}
\begin{center}
\includegraphics[width=3.0in]{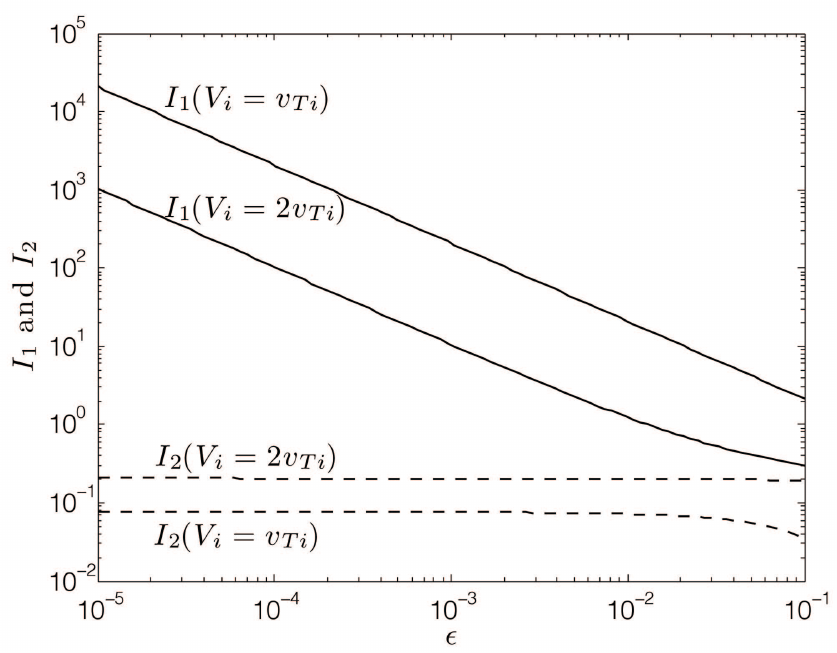}
\caption{Plots of $I_1$ and $I_2$ from Eqs.~(\ref{eq:i1}) and (\ref{eq:i2}) evaluated numerically excluding the interval $V_i/v_{Ti} \pm \epsilon$. The integral $I_2$ converges as $\epsilon \rightarrow 0$, but the integral $I_1$ diverges. Two values of the ion flow speed are shown for each integral: $V_i/v_{Ti} =1$ and $2$. }
\label{fg:i1i2}
\end{center}
\end{figure}

For a specific example, consider the predictions of Eqs.~(\ref{eq:kbc}) and (\ref{eq:kbcp}) in the two-fluid limit, where electrons and ions have Maxwellian distributions:  $f_e = n_e  \exp(-v^2/v_{Te}^2) /(\pi^{3/2} v_{Te}^3)$ and $f_i = n_i \exp [-(\vc{v} - \vc{V}_i)^2/v_{Ti}^2]/(\pi^{3/2} v_{Ti}^3) $ where $\vc{V}_i = V_i \hat{z}$. For these, Eqs.~(\ref{eq:kbc}) and (\ref{eq:kbcp}) reduce to
\begin{equation}
I_1 (V_i/v_{Ti}) =  \frac{1}{2\sqrt{\pi}} \int_{-\infty}^\infty dx \frac{e^{-x^2}}{(x-V_i/v_{Ti})^2} \leq \frac{T_i}{T_e} \label{eq:i1}
\end{equation}
and
\begin{equation}
I_2 (V_i/v_{Ti}) =  - \frac{1}{\sqrt{\pi}} \int_{-\infty}^\infty dx \frac{x e^{-x^2}}{x - V_i/v_{Ti}} \leq \frac{T_i}{T_e}  \label{eq:i2}
\end{equation}
respectively. We evaluate these integrals numerically by considering the Cauchy principal value: $\int_{-\infty}^\infty = \lim_{\epsilon \rightarrow 0} (\int_{-\infty}^{V_i/v_{Ti}-\epsilon} + \int_{V_i/v_{Ti} + \epsilon}^\infty )$. Figure~\ref{fg:i1i2}  shows that as $\epsilon$ gets small, the integral $I_2$ converges, but $I_1$ diverges. So, in the two-fluid limit, Eq.~(\ref{eq:kbc}) reduces to $\infty \leq T_i/T_e$. Riemann argues in section  4.1 of Ref.~\cite{riem:91} that Eq.~(\ref{eq:kbc}) can return fluid results by expanding $(v_z - V_i)^{-2}$ for $V_i \gg v_z$ in the integrand of Eq.~(\ref{eq:i1}). Indeed this does return Bohm's $V_i \geq c_s$ criterion. However, the result is fortuitous because the full integral actually diverges independent of the magnitude of $V_i$. Such an expansion neglects the low velocity ions argued to introduce important new physics in the kinetic formulation \cite{riem:12}, but which cause the integral to diverge. 

\begin{figure}
\begin{center}
\includegraphics[width=3.0in]{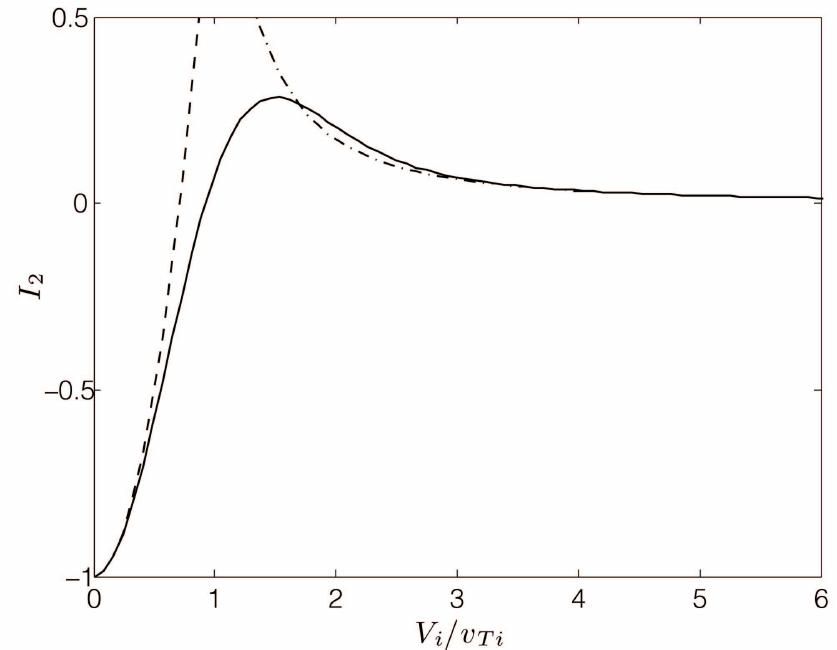}
\caption{The integral $I_2$ as a function of $V_i/v_{Ti}$ (solid line). Also shown are the small argument (dashed line) and large argument (dash-dotted line) asymptotic expansions.}
\label{fg:i2}
\end{center}
\end{figure}

Unlike $I_1$, the Cauchy principal value of the integral $I_2$ converges and has an analytic solution 
\begin{equation}
I_2 = -1 + \sqrt{\pi} (V_i/v_{Ti}) e^{-V_i^2/v_{Ti}^2} \textrm{erfi}(V_i/v_{Ti}),
\end{equation}
where $\textrm{erfi}$ is the imaginary error function (this is real for a real argument $V_i/v_{Ti}$). A plot of $I_2$ is shown in Fig.~(\ref{fg:i2}), along with the large argument expansion [$I_2 = v_{Ti}^2/(2 V_i^2) + 3 v_{Ti}^4/(4V_i^4)$] and small argument expansion ($I_2 = -1 + 2 V_i^2/v_{Ti}^2$). For $T_i/T_e \lesssim 0.285$, there are two solutions of Eq.~(\ref{eq:i2}). In the large argument limit, it reduces to Bohm's criterion $V_i \geq c_s$. In the small argument limit, it reduces to $V_i \leq v_{Ti} (1 + T_i/T_e)/\sqrt{2}$. The latter prediction is at odds with the fluid theory formulation, as well as experimental results. It is also noteworthy that for $T_i/T_e \gtrsim 0.285$, Eq.~(\ref{eq:i2}) is satisfied for any ion flow speed; there is little sheath-related information in this equation, as Riemann has shown previously from basic theoretical arguments \cite{riem:03,riem:06}. Thus, in the fluid limit, we find that the left side of Eq.~(\ref{eq:kbc}) diverges leading to an incorrect prediction. This divergence can be avoided by using the more primitive Eq.~(\ref{eq:kbcp}), but there is little sheath-related information in this equation. 

Reference~\cite{baal:11} considers another possible source of disagreement between experiments and Eq.~(\ref{eq:kbc}), related to errors associated with using non-Vlasov distributions in Eq.~(\ref{eq:kbc}). One way to justify the theory would be to include the collision terms in the analysis, then demonstrate that they are small in the final expression.\footnote{Here ``collision term'' refers to all terms on the right side of the kinetic equation, including ionization sources and sinks, collisions of charged particles with neutrals, Coulomb collisions, and wave-particle scattering.} As discussed in \cite{riem:12} and \cite{baal:11}, keeping collisions ($S$) leads to the additional term
\begin{equation}
\sum_s \frac{q_s}{E} \int_{-\infty}^\infty d^3v \frac{S}{v_z} \label{eq:source}
\end{equation}
on the right side of Eq.~(\ref{eq:kbc}). In his comment, Riemann points out that if $S$ is an even function of $v_z$, this term will vanish because the positive and negative velocity intervals will exactly cancel. In this way, the Vlasov formulation may be justified for symmetric collision terms. The seminal theories of Tonks and Langmuir \cite{tonk:29}, Emmert {\it et al.} \cite{emme:80} and Bissell and Johnson \cite{biss:87} are all two-scale theories that apply symmetric collision terms. These all lead to model distribution functions that satisfy Eq.~(\ref{eq:kbc}). Riemann also points out that in the limit $\lambda_D/l \rightarrow 0$, theories that obey Eq.~(\ref{eq:kbc}) can also be constructed using asymmetric collision terms because $E \rightarrow \infty$ at the sheath edge in this theoretical limit~\cite{vand:91}. 

Although theoretical models satisfying Eq.~(\ref{eq:kbc}) can be constructed based on symmetric collision terms, or asymmetric collision terms in the $\lambda_D/l \rightarrow 0$ limit, weakly collisional plasmas in the laboratory have both asymmetric collision terms and finite $\lambda_D/l$. The parity of ionization source terms is typically determined by the background neutral gas, which is often Maxwellian and leads to a symmetric source. However, sink terms are typically asymmetric in the laboratory frame. A frequently occurring example is ion charge exchange collisions where ions are removed from a flowing distribution. Another example is the sink associated with electron-neutral impact ionization. Near the sheath edge, the tail of the EVDF is often depleted in the direction facing away from the boundary because this population is lost to the boundary. Since tail electrons are responsible for the majority of ionization, this leads to an asymmetric sink term. Ion-electron Coulomb collisions also generate an asymmetric collision term because the ion flow is shifted in reference to the electron distribution. This is a ubiquitous example because the presheath electric field always drives flow, leading to a nonzero, and asymmetric, collision term.  Wave-particle scattering from flow-driven instabilities is another example that generates an asymmetric collision term~\cite{baal:09}. 

Because Eq.~(\ref{eq:kbc}) leads to divergent integrals and places undue importance on ions with small velocity in the $\hat{z}$ direction, we suggested in Ref.~\cite{baal:11} that the theory should be developed from positive exponent velocity moments instead. There is also a semantic reason for using positive exponent moments. Bohm's original criterion~\cite{bohm:49} is a condition pertaining to the ion fluid flow speed at the sheath edge: $V_i \geq c_s$. A kinetic generalization should pertain to a kinetic analog of the ion fluid flow speed. The $v_z^{-2}$ moment of $f_i$ does not have a clear connection to the fluid flow speed. It is well known that a hierarchy of fluid equations can be rigorously developed from positive exponent velocity moments of the kinetic equation. With this approach, the fluid flow velocity naturally arises through the moment expression $\vc{V}_s \equiv \int_{-\infty}^\infty d^3 v\, \vc{v} f_s/n_s$, where the species $s$ density is related to $f_s$ by $n_s \equiv \int_{-\infty}^\infty d^3 v\, f_s$. Reference \cite{baal:11} works from the philosophy that a kinetic generalization of Bohm's criterion should pertain to this kinetic analog of the fluid flow velocity. This approach provides an obvious physical correspondence between the kinetic and fluid theories, and simply returns the two-fluid results in the limit of thermodynamic equilibrium amongst individual species; the limit for which Eq.~(\ref{eq:kbc}) was just shown to diverge. 

Reference \cite{baal:11} shows that using the first two fluid moments of the full kinetic equation, along with Riemann's sheath criterion~\cite{riem:95}, provides the relation 
\begin{equation}
\sum_s q_s \biggl[ \frac{q_s n_s - \bigl( n_s\, dT_s/dz + d \Pi_{zz,s} / dz - R_{z,s} \bigr)/E}{ m_s V_{z,s}^2 - T_s} \biggr]_{z=0} \leq 0 .  \label{eq:longbohmcrit}
\end{equation}
Here density and flow velocity are defined in terms of the previously mentioned definitions, and the other definitions are the temperature $T_s \equiv  \int_{-\infty}^\infty d^3 v\, m_s v_r^2 f_s/(3n_s) =  m_s v_{Ts}^2/2$, stress tensor $\stres_s \equiv \int_{-\infty}^\infty d^3 v\, m_s ( \vc{v}_r \vc{v}_r -  v_r^2\, \ident /3 ) f_s$, and friction force density $\vc{R}_s \equiv \int_{-\infty}^\infty d^3 v\, m_s \vc{v} S(f_s)$. Here $\vc{v}_r = \vc{v} - \vc{V}_s$ and $S(f_s)$ can be any source, sink or collision term. Although Eq.~(\ref{eq:longbohmcrit}) has an obvious connection to fluid theory, it also has a kinetic interpretation and can be written explicitly in terms of the distribution functions by substituting the moment definitions for the fluid variables. One difficulty with this approach is that it depends on spatial derivatives of temperature and stress moments. Most of the time, but not always (e.g., see section 4.2.2 of \cite{baal:11}), the terms in parenthesis in Eq.~(\ref{eq:longbohmcrit}) are small because $E$ becomes large at the sheath edge. This is related to the $\lambda_D/l \ll 1$ ordering. In difficult cases, higher order moments and a closure scheme can be applied to deal with gradient terms. 

\begin{figure}
\begin{center}
\includegraphics[width=3.0in]{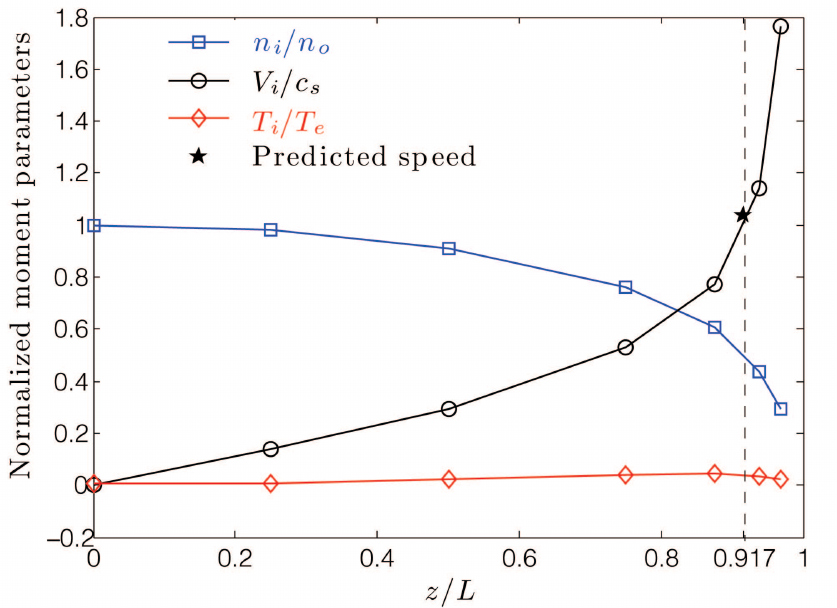}
\caption{Moment parameters calculated from the IVDFs shown in Fig.~\ref{fg:data}b for the $T_s = 0.01T_e$ case. The sheath edge was predicted by Sheridan \cite{sher:01} to be 0.917. The star shows the predicted speed at the sheath edge using Eq.~(\ref{eq:longbohmcrit}).}
\label{fg:mom}
\end{center}
\end{figure} 

Finally, we use the numerical data from Fig.~\ref{fg:data}b to test the two criteria from Eqs.~(\ref{eq:kbc}) and (\ref{eq:longbohmcrit}). Considering the $T_s = 0.01 T_e$ data from Sheridan's simulations, Fig.~\ref{fg:mom} shows the density, flow speed and temperature calculated from the appropriate moments of the IVDFs shown in Fig.~\ref{fg:data}b. For this situation (with Boltzmann electrons), Eq.~(\ref{eq:longbohmcrit}) reduces to $V_i \geq \sqrt{c_s^2 + v_{Ti}^2/2}$, leading to the prediction $V_i = c_s \sqrt{1+T_e/T_i} \simeq 1.04 c_s$ at the sheath edge. This prediction is represented by the star in the figure where the sheath edge, as determined from cold Tonks-Langmuir theory by Sheridan, is predicted to be $x/L = 0.917$. As Fig.~\ref{fg:mom} shows, this prediction agrees well with the numerical results. Conversely, if we try to apply this data to Eq.~(\ref{eq:kbc}), we find that $\int d^3v\, f_i/v_z^2 \rightarrow \infty$. Taking $f_e$ to be Maxwellian, which is consistent with the Boltzmann relation used in the simulations, the right side is $n_e/T_e$. Thus, Eq.~(\ref{eq:kbc}) provides the criterion $\infty \leq n_e/T_e$, which is not consistent with the data. Applying the same analysis to the other data in Fig.~\ref{fg:data} similarly leads to divergences in Eq.~(\ref{eq:kbc}), but the fluid moment approach predicts $V_i \gtrsim c_s$, in agreement with the data.  

In conclusion, the conventional KBC places undue importance on low velocity ions, which leads to divergences when the ion distribution function has particles at $v_z =0$.  We find that it is more productive to work from the perspective of generalizing the fluid theory to account for arbitrary distributions through the positive-exponent velocity moments that are usually used to define the fluid variables. This avoids the problem of divergent integrals. It also provides an obvious connection to the original fluid criterion, and gives predictions that agree with experimental and numerical data.

\ack 
The authors thank T.\ Sheridan, G.\ Severn and W.\ Miloch for permission to reprint figures from their previously published work, and for providing the data we used to test the theories. This work was carried out under the auspices of the National Nuclear Security Administration of the U.S. Department of Energy at Los Alamos National Laboratory under Contract No. DE-AC52-06NA25396, and in part from DOE through grant no. DE-FG02-8653218.

\end{document}